\documentclass[fleqn,usenatbib]{mnras}
\usepackage{txfonts}

\usepackage[T1]{fontenc}
\usepackage{ae,aecompl, longtable, lscape}

\usepackage{color, soul, ulem,graphicx, amsbsy}

\newcommand{\sw}{_{\rm sw}}

\title[Stellar wind effects on the atmospheres of close-in giants]{Stellar wind effects on the atmospheres of close-in giants: a possible reduction in escape instead of increased erosion}
\author[Vidotto \& Cleary]{A. A. Vidotto\thanks{E-mail: aline.vidotto@tcd.ie}, A.~Cleary \\
School of Physics, Trinity College Dublin, the University of Dublin, Dublin-2, Ireland
}

\date{Accepted XXX. Received YYY; in original form ZZZ}

\pubyear{2020}

\begin{document}
\label{firstpage}
\pagerange{\pageref{firstpage}--\pageref{lastpage}}
\maketitle

%%%%%%%%%%%%%%%%%%%%%%%%%%%%%%%%%%%%%%%%%%%%%%%%%%%%%%%%%%%
\begin{abstract}
The atmospheres of highly irradiated exoplanets are observed to undergo hydrodynamic escape. However, due to strong pressures, stellar winds can confine planetary atmospheres, reducing their escape. Here, we investigate under which conditions atmospheric escape of close-in giants could be confined by the large pressure of their host star's winds. For that, we simulate  escape in planets at a range of orbital distances ([0.04, 0.14]~au), planetary gravities ([36\%, 87\%] of Jupiter's gravity), and ages ([1, 6.9]~Gyr). For each of these simulations, we calculate the ram pressure of these escaping atmospheres and compare them to the expected stellar wind external pressure to determine whether a given atmosphere is confined or not. We show that, although younger close-in giants should experience higher levels of atmospheric escape, due to higher stellar irradiation, stellar winds are also stronger at young ages, potentially reducing escape of young exoplanets. Regardless of the age, we also find that there is always a region in our parameter space where atmospheric escape is confined, preferably occurring at higher planetary gravities and  orbital distances. We investigate confinement of some known exoplanets and find that the atmosphere of several of them, including $\pi$ Men c, should be confined by the winds of their host stars, thus potentially preventing escape in highly irradiated planets. Thus, the lack of hydrogen escape recently reported for $\pi$ Men c could be caused by the stellar wind. 
\end{abstract}
%%%%%%%%%%%%%%%%%%%%%%%%%%%%%%%%%%%%%%%%%%%%%%%%%%%%%%%%%%%
\begin{keywords}
stars: planetary systems --  stars: winds, outflows -- planet-star interactions 
\end{keywords}

%%%%%%%%%%%%%%%%%%%%%%%%%%%%%%%%%%%%%%%%%%%%%%%%%%%%%%%%%%%%
%
\section{Introduction}\label{sec.intro}
Mass loss plays a key role during the lifetime of exoplanets, influencing their potential to develop and host life \citep{2009A&ARv..17..181L}. When planets lose mass, their orbital evolution changes, which may lead to planetary engulfment \citep{2016A&A...593A.128P}. Planetary mass loss also regulates angular momentum evolution \citep{2014ApJ...788..161T}, which is deeply connected to magnetic field generation \citep{2012Icar..217...88Z}, which in turn may affect atmospheric retention. As a consequence, evolution of planetary mass loss is also crucial for understanding planet populations \citep{2009MNRAS.396.1012D}. 

Close-in exoplanets are bathed in the intense irradiation and outflows of their host stars, both of which can affect planetary mass loss. Irradiation heats the atmospheres, causing them to inflate and more likely to outflow through a hydrodynamic escape mechanism. While the effects of irradiation on the mass-loss process of exoplanets have been largely studied \citep{2003ApJ...598L.121L, 2004A&A...419L..13B, 2004A&A...418L...1L, 2008A&A...477..309P, 2011A&A...529A.136E,2004Icar..170..167Y, 2005ApJ...621.1049T, 2007P&SS...55.1426G, 2009ApJ...693...23M}, and several works have studied how stellar winds interact with escaping atmospheres \citep{2007ApJ...671L..57S, 2012ApJ...744...70K,2015ApJ...813...50K, 2014ApJ...795..132S, 2014MNRAS.438.1654V, 2018MNRAS.479.3115V, 2015ApJ...808..173T,2017MNRAS.466.2458C,2018MNRAS.481.5296V, 2019MNRAS.483.1481D, 2019MNRAS.483.2600D,2019MNRAS.487.5788E,2019ApJ...873...89M}, the effects of stellar ejecta in {\it confining} the escape of exoplanetary atmospheres have been less explored. This is the subject of the present paper. 

Planetary outflows can be shaped or even completely transformed through the interactions with stellar outflows, which are comprised of quiescent {stellar winds} and violent, short releases of {coronal mass ejections}. As such stellar outflows propagate through the interplanetary medium, they interact with any orbiting exoplanet and cause a pressure confinement around (otherwise freely) expanding atmospheres of exoplanets. Here, we focus on the case of quiescent stellar winds. Given the extreme stellar wind conditions around close-in exoplanets \citep{2009ApJ...699..441V,2010ApJ...720.1262V,2011MNRAS.412..351V, 2012MNRAS.423.3285V,2014MNRAS.438.1162V,2015MNRAS.449.4117V}, the atmospheres of these planets are under strong pressure confinement. This implies that the high external local pressure of stellar winds can have a profound effect on exoplanetary mass loss and can even suppress planetary outflows \citep{2011ApJ...730...27A,2011ApJ...728..152T,2014MNRAS.444.3761O}.  Such an extreme condition does not exist in the solar system.

\begin{figure*}
	\includegraphics[width=.4\textwidth]{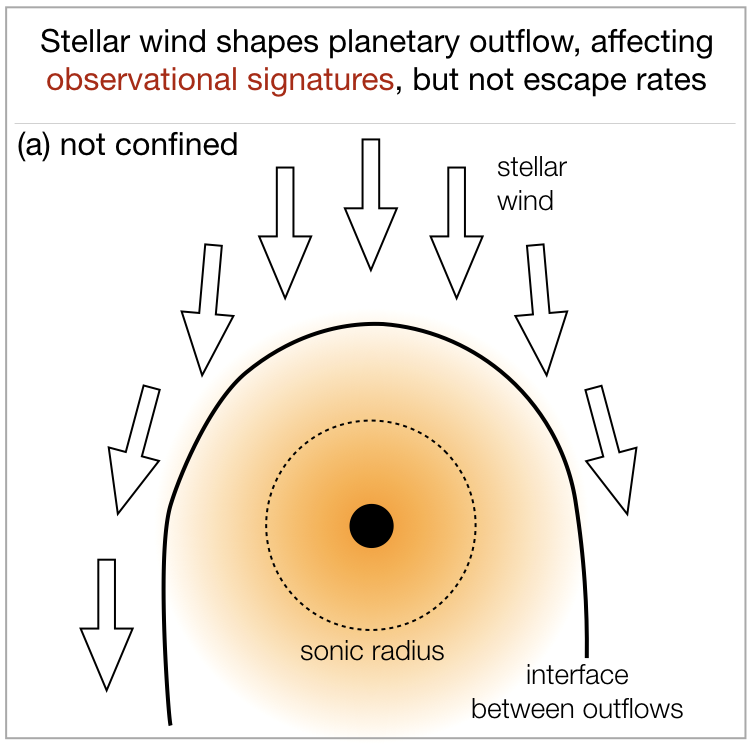}
	\includegraphics[width=.4\textwidth]{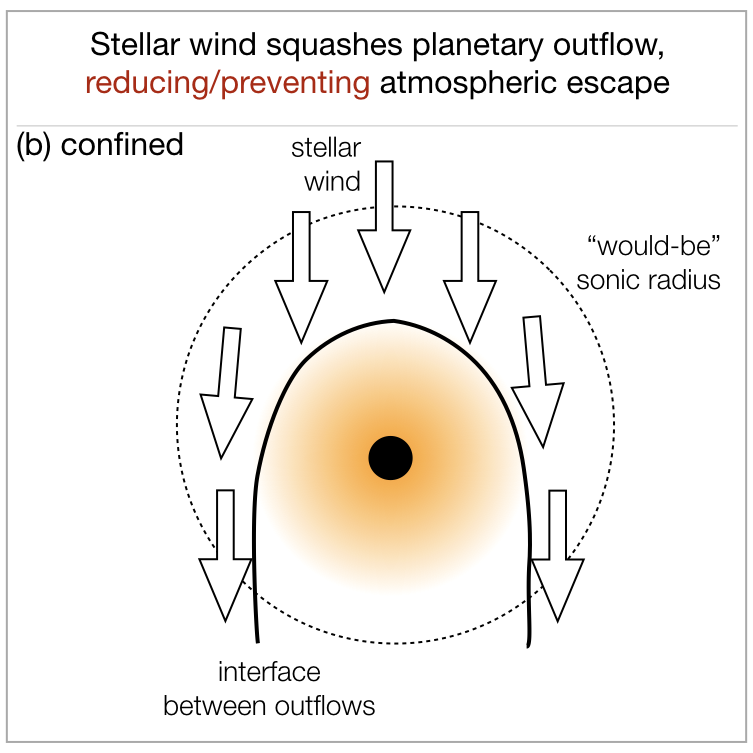}
    \caption{Sketch of how stellar outflows (arrows coming from above) can affect the structure of escaping planetary atmospheres (orange region), depending on the relative position of the sonic radius (dashed) with respect to the interface boundary between stellar and planetary outflows (thick black curve). The interface boundary is calculated using balance of ram pressures from the stellar wind and planetary outflow. Mathematically, the regimes on panel a (not-confined) and panel b (confined) can be described by Equations (\ref{eq.non_confinement}) and (\ref{eq.confinement}), respectively. This figure is a rough representation of flow-flow interactions. The precise geometry of the interaction depends, among others, on the strength of stellar winds, which might include formation of accretion streams \citep[e.g.,][]{2015A&A...578A...6M,2016ApJ...832..173S,2019ApJ...873...89M}.}
    \label{fig.pl_conditions}
\end{figure*}

The key concept for determining whether  the stellar wind simply shapes the escaping atmosphere or whether the stellar wind affects the entire structure of the planetary atmosphere is sketched in Figure~\ref{fig.pl_conditions}. Similar to stellar wind theory, the physical solution for planetary outflows has `critical radii' (sonic, Alfven, etc) that determine outflow properties \citep{1958ApJ...128..664P, 1967ApJ...148..217W}. Here, we only consider the case of unmagnetised planets, so the critical radii is known as sonic radius, which is the radius where the velocity of the escaping planetary outflow reaches sonic speed. In Figure~\ref{fig.pl_conditions}, the stellar wind is shown by the arrows coming from the top. The wind then interacts with the escaping planetary atmosphere, shown in orange, forming an interface boundary (thick black curve). We will discuss later that this interface can represent, for example, a shock transition. On first-order approximation, the location of the interface is given by pressure balance between the stellar wind and the evaporating atmosphere.

An escaping atmosphere accelerates from low velocities at the lower base upwards, until it reaches the boundary with the stellar wind. In doing that, the atmospheric outflow crosses the sonic radius (dashed circles), i.e., it goes from subsonic velocities to supersonic velocities. In Figure~\ref{fig.pl_conditions}a, the sonic  radius of the escaping planetary atmosphere is inside the interface that separates the stellar wind from the planetary outflow.  Because the interaction happens at supersonic velocities, information cannot be passed back to the base of the planetary atmosphere. In this case, the inner layers of the planetary outflow structure cannot be affected by the stellar wind. Nevertheless, the stellar wind delineates/shapes the escaping atmosphere, affecting the escaping signature that could be detected with transmission spectroscopy (e.g., asymmetric line profiles). 

On the contrary, in Figure~\ref{fig.pl_conditions}b, the interface separating stellar wind and planetary outflow is deep enough so that the stellar wind interacts with the still subsonic planetary outflow. In this regime, the atmospheric structure is altered by the stellar wind, which can reduce or prevent escape rates \citep{2009ApJ...693...23M,2011ApJ...730...27A}. Multi-dimensional simulations carried out by \citet{2016ApJ...820....3C} showed that the planetary outflow can be ``shut off when the stellar wind penetrates inside where the sonic point would have been.'' This suggests that active and moderately-active stars, with their intense winds and magnetism, could actually reduce or even suppress mass loss from their exoplanets, challenging the commonly accepted scenario that evaporation rates are much higher in close-in planets or planets orbiting young (i.e., more magnetically active) stars. This means that atmospheres of planets embedded in strong stellar winds might not necessarily be more prone to erosion. 

Here, we conduct a parametric study to delineate the parameter space where stellar winds have the largest effects on  mass loss of close-in giants. Here, we only consider the cases of unmagnetised outflows. This paper is divided as follows. The stellar wind models and planetary escape model are shown in Section \ref{sec.model}. Section \ref{sec.confinement} illustrates the  condition required for confinement of escaping planetary atmospheres by the winds of their host stars. Our parametric study is presented in Section \ref{sec.parametric_results} and, in Section \ref{sec.discussion}, we present further discussion of our model. Section \ref{sec.conclusion} shows our final concluding remarks. 

The novelty of our work is that we investigate when stellar winds have the largest effects on  mass loss of close-in giants, by focusing on the nature of the planetary outflow being subsonic/supersonic. Previous works studying the interaction between close-in planets and stellar winds have focused on cases where the planetary outflow is supersonic \citep[e.g.][]{2018MNRAS.479.3115V, 2019MNRAS.483.1481D, 2019MNRAS.483.2600D,2019ApJ...873...89M}. Alternatively, previous works have explored the effects of a sub/super-critical (i.e., sub/super-Alfvenic, sub/supersonic) stellar wind on exoplanets \citep[e.g.][]{2010ApJ...722L.168V, 2013A&A...552A.119S, 2013A&A...557A..31L, 2019ApJ...881..136S, 2019ApJ...872..113F, 2019arXiv191203736F}, which is relevant for studies of star-planet interactions and their possible signatures \citep{2004ApJ...602L..53I,2009A&A...505..339L}.

%%%%%%%%%%%%%%%%%%%%%%%%%%%%%%%%%%%%%%%%%%%%%%%%%%%%%%%%%%
\section{Stellar winds and planetary escape models}\label{sec.model}
In this work, we use two different models to compute the physical conditions of the stellar wind profile and of a planetary outflow. As we will describe below, both models adopt a fluid description and they share many similarities. This means that many lessons from stellar wind theory can be applied to the study of planetary outflows.  Some key differences, though, exist between the planetary and stellar outflows, mainly on their energetics and ionisation state. 

Winds of low-mass stars are thought to be heated by magnetohydrodynamic waves \citep[e.g.,][]{2008ASPC..384..317C,2013PASJ...65...98S}. The precise process is currently now well understood, but it is believed that they are of magnetic origin (e.g., dissipation of Alfven waves, magnetic reconnection). Winds of low-mass stars are expected to reach high coronal temperatures ($\gtrsim 10^6$~K), leading to a fully ionised outflow. Once the flow reaches such high temperature, the wind becomes thermally driven, i.e., pressure gradient is the main force acting against the gravitational potential of the star. 

Outflows from close-in exoplanets, on the other hand, are primarily heated by  stellar EUV irradiation. This is substantially different from stellar winds: in the stellar wind case, the heating is injected from the lower boundary (e.g., from stellar photosphere or the corona) either by injecting a flux of waves at the wind base or  by assuming that the wind base is already at typical coronal temperatures. In the planetary outflow case, the heating comes from an outer boundary, i.e., from stellar photons that enter the outer boundary of the simulation domain. As the photons penetrate through the planetary atmosphere, they deposit energy. This heating deposition leads to an increase in temperature from the base of the atmosphere and reach a peak temperature of $\sim 10^4$~K, decreasing beyond that (mostly due to adiabatic expansion). Outflows from close-in giants are partially neutral, contrary to fully ionised stellar winds. 

In this section, we present the models used to compute the physical conditions of stellar winds and exoplanetary outflows.

%%%%%%%%%%%%%%%%%%%%%%%%%%%%%%%%%%%%%%%%%%%%%%%%%%%%%%%%%%
\subsection{Planetary outflow model: hydrodynamic escape}
Several mechanisms can drive planetary mass loss: non-thermal escape, Jeans escape and (magneto)hydrodynamic escape, for instance. In the case of close-in planets, the atmospheric temperatures are sufficiently high to create expanded, high-density atmospheres, which means that the gas is collisional and can be treated as a fluid. Due to these high densities, the hydrodynamical escape is the evaporation process that can produce the largest escape rate in close-in exoplanets. It is important at high incident fluxes -- it takes place at close-in planets and/or planets orbiting active stars, when the incident extreme UV (EUV) fluxes are several orders of magnitude larger than the EUV solar flux at Earth. 

 Here, we use the model from \citet{2019MNRAS.490.3760A} to compute the hydrodynamic properties of  escaping atmospheres of close-in giants. This model is based on \citet{2009ApJ...693...23M},  and it considers planetary gravity, stellar tidal forces, photoionisation heating from stellar EUV radiation, cooling from  Ly-$\alpha$ radiation and ionisation balance. Assuming a steady state outflow and spherical symmetry, the mass conservation equation becomes
\begin{equation}\label{eq.mdot}
\frac{d (r^2 \rho {u}) }{d r } = 0  ,
\end{equation}
where $r$ is the radial coordinate from the centre of the planet, $\rho$ and $u$ are the  mass density and velocity, respectively. Integrating Equation (\ref{eq.mdot}) over the area, we have that the evaporation rate of the escaping atmosphere is $\dot{M} = 4\pi r^2 \rho {u} $. Because we assume that atmospheric escape takes place in the entire surface of the planet, neglecting the fact that, for example, in the night side the mass flux might be reduced, 1D models can overestimate escape rates \citep[e.g.,][]{2009ApJ...693...23M, 2015ApJ...815L..12J}.
 
 The forces acting on the planetary outflow are the thermal pressure gradient, gravitational force and tidal force (i.e., the sum of the centrifugal force and differential stellar gravity along the ray between the planet and star, e.g., \citealt{2007P&SS...55.1426G}). Thus, the momentum equation becomes 
\begin{equation}\label{eq.momentum}
{u}\frac{d {u}}{d r } = -\frac{1}{\rho}\frac{d P}{d r}   -    \frac{G M_{ \rm pl}}{r^2}+\frac{3GM_*r}{a_{\rm orb}^3}  ,
\end{equation}
where  $P$ is the thermal pressure, $G$ the gravitational constant, $M_{\rm pl}$ and $M_*$ are the masses of the planet and of the star, respectively, and $a_{\rm orb}$ is the orbital distance. We assume an ideal gas, thus $P=\rho k_B T/m$, where $k_B$ is the Boltzmann constant, $T$ the temperature, and $m$ is the mean molecular weight. Here, we assume a hydrogen plasma, hence, $m = (n_e m_e + n_p m_p + n_n m_p)/(n_e+n_p+n_n) \simeq (n_p + n_n)m_p/(2n_p + n_n) $ where $n_e$, $n_p$ and $n_n$ are the electron density, the proton density and the density of neutral hydrogen, respectively, and $m_e$ and $m_p$ are the  electron and proton masses, respectively.   The energy equation is given by
\begin{equation}\label{eq.energy}
    \rho {u} \frac{d}{d r} \left[\frac{k_BT}{(\gamma - 1)m} \right] = \frac{k_BT}{m}{u} \frac{d\rho}{d r }+Q-C  ,
\end{equation}
where  $Q$ and $C$ are the volumetric heating and cooling rates, respectively. We assume that the escaping atmospheric gas cools by radiative losses resulting from collisional excitation
$$ C= 7.5 \times 10^{-19} n_p n_n \exp[{-1.183\times 10^5}/{T}].$$ 
Given that the heating is generated by the photoionisation of hydrogen, the volumetric heating rate is given by
$$Q=\epsilon F_{\rm EUV}e^{-\tau} \sigma_{\nu_0}n_{n},$$
where $F_{\rm EUV}$ is the stellar EUV flux received at the orbit of the planet, $\tau$  the optical depth to ionising photons and $\sigma_{\nu_0}$ is the cross section for the photoionisation of hydrogen  given by $\sigma_{\nu_0}=6\times 10^{-18} \left( {e_{\rm in}}/{13.6\, {\rm eV}}\right)^{-3}=1.89 \times 10^{-18}{\rm cm}^2$ \citep{spit}. Here, we assume monochromatic flux with energy $e_{\rm in}=20$~eV. Additionally, we also assume that only a fraction $\epsilon= (e_{\rm in}- {13.6 \,\rm{eV}})/{e_{\rm in}}=32\%$ of $F_{\rm EUV}$ is converted to thermal energy to heat the atmosphere.

The proton number density, which is the same as the electron number density, is calculated assuming ionisation balance
 \begin{equation}\label{eq: ion_bal}
 \frac{1}{r^2}\frac{d}{d r}(r^2n_p{u}) =
\frac{n_n F_{\rm EUV}e^{-\tau}\sigma_{\nu_0}}{e_{\rm in}}- n_p^2\alpha_{\rm rec}
\end{equation}
where the terms on the right-hand-side are the volumetric rate of photoionisations and of radiative recombinations [cm$^{-3}$\,s$^{-1}$], respectively. Here, $\alpha_{\rm rec} = 2.7 \times 10^{-13} (T/10^4)^{-0.9}$ cm$^{3}$/s  is the case B radiative recombination coefficient for hydrogen ions \citep{1995MNRAS.272...41S, 2006agna.book.....O}. The number density of neutral hydrogen is related to total number density $n=\rho/m$ by $n= 2 n_p + n_n$. 

Equations (\ref{eq.mdot}) to (\ref{eq: ion_bal}) form a system of coupled differential equations. Our calculations start at the planetary radius $r=R_{\rm pl}$ and extend out to $r=R_{\rm Roche} = a_{\rm orb}[M_{\rm pl}/(3M_\star)]^{1/3}$, i.e., the point where the stellar gravitational force balances the planetary gravitational force. The only physical solution of this system of equations is the one in which the velocity increases from $u (R_{\rm pl})\simeq 0$ at the base of the planetary outflow, passes through the sonic point at $r=r_s$, where $u(r_s)=(\gamma k_B T/m)^{1/2}$, and becomes supersonic beyond that point, similarly to stellar wind theory. There is no analytical solution for knowing a priori the location of the sonic point, the velocity profile, etc. Thus, we use an iterative shooting method and check convergence of the solution by ensuring that density and velocity profiles do not vary by more than 1\% between two  iterations. We also calculate the Knudsen number $K_n$  to check that the atmosphere remains collisional ($K_n \ll 1$). 

The mass, radius and orbital distance of the planet are input parameters of our models, that we vary in our parametric study (details will be presented in Section \ref{sec.parametric_results}). Throughout this work, the mass of the star is assumed to be $1M_\odot$.  
Another input in our planetary wind models is the EUV luminosity $L_{\rm EUV}$ of the star, which is used to derive the input energy flux, $F_{\rm EUV} = L_{\rm EUV}/(4\pi a_{\rm orb}^2)$. When stars are younger, they are more magnetically active and thus have higher $L_{\rm EUV}$. In this work, we adopt 4 different ages for the systems studied, ranging from 1~Gyr to 6.9~Gyr. For each of these ages, we use the age--$L_{\rm EUV}$ relation from \citet{2015A&A...577L...3T}, to derive the values of $L_{\rm EUV}$, which are presented in Table \ref{tab.inputs}. All these aforementioned input parameters are set either by observations or by the physical characteristics of the systems we wish to study. Nevertheless, our models also have free parameters, namely the temperature and density (or pressure) at the base of the evaporating atmosphere. Here, we set the base at $r=R_{\rm pl}$ and assume the temperature there is 1000~K and the density is $4\times 10^{-13}$~g~cm$^{-3}$. These are similar to values used in \citet{2009ApJ...693...23M, 2019MNRAS.490.3760A}. We remind however that \citet{2009ApJ...693...23M} demonstrated that changing these values had negligible effect on the physical properties of the atmospheric escape.

\begin{table}
\caption{Input parameters for the simulations of planetary winds and stellar winds. Superscripts A and B indicate different stellar wind models (Equations (\ref{eq.baseT_A}) and  (\ref{eq.baseT_B})).}\label{tab.inputs}
\begin{tabular}{ccccccccc}
\hline 
age & $ L_{\rm EUV}$  & $P_{\rm rot, \star} $  &  ${n\sw}_0$  & ${T\sw}_0^A$ &   ${T\sw}_0^B$\\ 
(Gyr)&$ (10^{-6} L_\odot$) &(d) & ($10^{8}~ \rm{cm}^{-3}$) & (MK)  & (MK) \\\hline \hline
1.0 & $ 51$ & $12$ &    $1.6$ & $2.64$ & $1.62$ \\
2.7	& $ 9.3$ & $21$ & $1.2$ & $2.06$ & $1.54$\\
4.6	& $ 4.6$ & $27$ & $1.0$ & $1.50$ & $1.50$ \\
6.9 &  $3.3$ & $33$ &   $0.9$ & $1.17$ & $1.47$\\ \hline
\end{tabular}
\end{table}

%%%%%%%%%%%%%%%%%%%%%%%%%%%%%%%%%%%%%%%%%%%%%%%%%%%%%%%
\subsection{Stellar wind models}
There are many levels of complexity in which one could model the winds of low-mass stars.\footnote{{In this work, we use the fluid description to describe stellar winds, which implicitly assumes that  winds of low-mass stars are collisional. However, if these winds are similar to the solar wind, this is likely not the case. \citet{2011SGeo...32....1E} provided a broad overview of the several-decade discussion between kinetic and fluid treatments of the solar wind. As they put it, the fluid model can be seen as a ``global'' or macroscopic description of the solar wind, while the kinetic model are more suited to describe specific processes that cannot be described using fluids. The two approaches are complementary to each other, and both have their advantages and disadvantages. One particular advantage of the fluid description is its easier numerical implementation.}}  Because these stars are magnetised and rotating, their winds can be treated as `magnetic rotator' winds. Spectropolarimetric observations have revealed that these stars can harbour complex large-scale magnetic fields, which also affect the structure of their winds. For these reasons, 3D models better capture the complex structure of winds of low-mass stars \citep[e.g.][]{2015MNRAS.449.4117V}. However, the drawback is that these models take long to run and are thus not ideal for large parametric studies. Aiming for a more efficient wind calculation, here we opt to model the winds of low-mass stars using a polytropic wind model. For all the winds simulated here, we assume a star similar to the sun, with mass $M_\star = 1 M_\odot$ and radius $R_\star = 1 R_\odot$. A polytropic wind means that the density and pressure are related, such that $P\sw \propto \rho\sw ^\Gamma$, where $\Gamma$ is the polytropic index. In practice, the polytropic wind mimics energy deposition processes, without solving for a more complex energy equation. For example, polytropic indices near 1 means that the wind is nearly isothermal. 

The hydrodynamic equations describing a polytropic wind are the mass conservation equation
\begin{equation}\label{eq.mdot_1.sw}
\frac{d (R^2 \rho\sw {u\sw}) }{d R } = 0 ,  
\end{equation}
and the momentum equation
\begin{equation}\label{eq.momentum.sw}
{u\sw}\frac{d {u\sw}}{d R } = -\frac{1}{\rho\sw}\frac{d P\sw}{d R}   -    \frac{G M_{\star}}{R^2}  ,
\end{equation}
where $R$ is the radial coordinate from the centre of the star, $\rho\sw$, $u\sw$ and $P\sw$ are the  mass density, velocity and thermal pressure of the stellar wind, respectively. We again assume an ideal gas, thus $P\sw=\rho\sw k_B T\sw/m\sw$. Here, we assume a fully ionised hydrogen wind, which implies that the mean mass of the wind particle is $m\sw=0.5 m_p$. The two previous equations assume that the stellar wind is in steady state and spherically symmetric.  Thus, integrating Equation (\ref{eq.mdot_1.sw}) over the area, we can derive the mass-loss rate of a stellar wind
\begin{equation}\label{eq.mdot.sw}
\dot{M}\sw = 4\pi R^2 \rho\sw {u\sw} \, , 
\end{equation}
which is a constant of the wind. The forces acting on a polytropic wind [Equation  (\ref{eq.momentum.sw})] are the gradient of the thermal pressure and the stellar gravity force. There are infinite solutions for the momentum equation, but, once again, the only physical solution is the transonic one, passing through the critical point $R=R_c$, i.e., when $u\sw (R_c) = (\Gamma k_B T\sw/ m\sw)^{1/2}$. To find this solution, we use a shooting method.

The free parameters of our stellar wind model are the base temperature ${T\sw}_0$, the number base density ${n\sw}_0 ={\rho\sw}_0/m\sw$ and $\Gamma$. We assume $\Gamma = 1.05$ in our models, as values in the range of 1.05 -- 1.15 are often adopted in wind simulations \citep{2018MNRAS.481.5296V}, in accordance to observed values at closer distances to the sun being around 1.1 \citep{2011ApJ...727L..32V}. We use two different empirical results to define ${n\sw}_0$ and ${T\sw}_0$, both of which correlate these quantities to the rotation of the star (which is here used as a proxy for age, as we will see later). Given our different assumptions, these two different models are called A and B. For these two models, the base density decreases for increasing rotation periods $P_{\rm rot, \star}$ (i.e., towards slower rotators) as
\begin{equation}\label{eq.basedens_original}
    {n\sw}_0 = 10^8 \left(\frac{P_{\rm rot, \odot}}{P_{\rm rot, \star} }\right)^{0.6} {\rm cm}^{-3}, 
\end{equation}
where $P_{\rm rot, \odot} = 27.2$~days is the rotation period of the present-day sun. 
The $0.6$ exponent is derived from X-ray observations \citep{2003ApJ...599..516I} and has been adopted in a series of wind models \citep{2007A&A...463...11H, 2016ApJ...832..145R,2018MNRAS.476.2465O,2019MNRAS.489.5784C}. Here, we scale the base density to match the present-day solar wind base density of $10^8$~cm$^{-3}$ in Equation (\ref{eq.basedens_original}). Given that the previous equation is given in terms of rotation period, to assign an age for our models, we use a Skumanich law \citep{1972ApJ...171..565S}
\begin{equation}\label{eq.skumanich}
P_{\rm rot, \star} = \left( \frac{\rm age}{4.6~ \rm Gyr} \right)^{1/2} P_{\rm rot, \odot } \, .
\end{equation}
Equation (\ref{eq.skumanich}) is a reasonable approximation for stars older than about 1~Gyr, but should not be adopted for younger stars. Subbing Equation (\ref{eq.skumanich}) in Equation (\ref{eq.basedens_original}), we have
\begin{equation}\label{eq.basedens}
    {n\sw}_0 = 10^8 \left( \frac{4.6~ \rm Gyr}{\rm age} \right)^{0.3} {\rm cm}^{-3}. 
\end{equation}

For the base temperature, we use two different empirical scalings 
\begin{description} 
\item \underline{Model A:} 
\begin{eqnarray}
  {T\sw}_0= 1.5 \left(\frac{P_{\rm rot, \odot}}{P_{\rm rot, \star} }\right)^{1.2} {\rm MK, ~ for} ~ P_{\rm rot, \star} \gtrsim 0.7 P_{\rm rot, \odot} ,   \nonumber \\
    {T\sw}_0 =  1.98\left( \frac{P_{\rm rot, \odot}}{P_{\rm rot, \star} }\right)^{0.37} {\rm MK, ~for} ~ P_{\rm rot, \star} \lesssim 0.7 P_{\rm rot, \odot} .   \label{eq.baseT_A_original} 
\end{eqnarray}
\item \underline{Model B:} 
\begin{equation}
  {T\sw}_0= 1.5 \left(\frac{P_{\rm rot, \odot}}{P_{\rm rot, \star} }\right)^{0.1} {\rm MK}.  \label{eq.baseT_B_original}
\end{equation}
\end{description} 
In Model A, Equation (\ref{eq.baseT_A_original}) is a broken power law derived from X-ray data \citep{2015A&A...578A.129J} by \citet{2018MNRAS.476.2465O}. In Model B, the slope $0.1$ in Equation (\ref{eq.baseT_B_original})   is the reference case derived by \citet{2007A&A...463...11H}. Note that Models A and B both provide $  {T\sw}_0= 1.5$~MK at the solar rotation period. This constant is set so that either model is able to reproduce the present-day solar wind mass-loss rate. We can rewrite the two previous equations using Equation (\ref{eq.skumanich}), thus, 
\begin{description} 
\item \underline{Model A:} 
\begin{eqnarray}
  {T\sw}_0= 1.5 \left( \frac{4.6~ \rm Gyr}{\rm age} \right)^{0.6} {\rm MK, ~ for} ~ {\rm age} \gtrsim 2.2 {\rm ~Gyr} ,   \nonumber \\
    {T\sw}_0 =  1.98  \left( \frac{4.6~ \rm Gyr}{\rm age} \right)^{0.19} {\rm MK, ~for} ~ {\rm age} \lesssim 2.2 {\rm ~Gyr}  .   \label{eq.baseT_A} 
\end{eqnarray}
\item \underline{Model B:} 
\begin{equation}
  {T\sw}_0= 1.5  \left( \frac{4.6~ \rm Gyr}{\rm age} \right)^{0.05} {\rm MK}.  \label{eq.baseT_B}
\end{equation}
\end{description} 

The input parameters for the stellar wind models are also shown in Table \ref{tab.inputs}. In our simulations, only the 1~Gyr case for Model A falls in the `fast rotating branch', with $P_{\rm rot, \star} \lesssim 0.7 P_{\rm rot, \odot}$.

%%%%%%%%%%%%%%%%%%%%%%%%%%%%%%%%%%%%%%%%%%%%%%%%%%%%%%%%%
\section{Condition for stellar wind confinement of planetary atmospheres}\label{sec.confinement}
In deriving the transonic hydrodynamic equations for stellar winds, there is an implicit result that, at the outer boundary, at very large distances ($R \to \infty$), the thermal pressure of the flow goes to zero. In reality, this is not entirely correct. For example, the solar wind does not expand into vacuum, but it is actually bound by the interstellar medium (ISM). However, because this interaction happens way beyond the critical points of the solar wind, assuming that  $P_{\rm sw}(R\to \infty)\simeq 0$ is not unreasonable. Eventually, a terminal shock transition between the solar wind and the ISM develops. The position of the termination shock in the heliosphere is $\sim 80$ au, while the sonic point of the solar wind is within a fraction of an au. 
 
Imagine that we can now change the pressure of the ISM: as the external ambient pressure  increases, the  shock moves closer and closer to the star. In the limit where the external pressure is high enough that the shock position happens exactly at the sonic point, the flow would never reach a supersonic solution: the flow remains a subsonic `breeze'. It has been demonstrated through numerical simulations that if one increases the external ambient pressure even further, the flow reverses, thus collapsing the wind to an accretion inflow \citep{1989A&A...226..209K, 1998A&A...330L..13D}.

Given that we use a similar theory to describe atmospheric escape in hot Jupiters, it is natural to question whether it is valid to assume a near `vacuum expansion' for a planetary outflow. Similarly to stellar winds being bounded by the ISM, planetary hydrodynamic outflows are bounded by the winds of their host stars, which can exert a significant, non-zero pressure on what would otherwise be a freely expanding atmosphere. In reality, stellar winds can have a high external local pressure, especially in the case of close-in planets and/or planets orbiting more active stars \citep{2013A&A...557A..67V, 2015MNRAS.449.4117V}. 

This means that a planetary outflow might not become a `wind' and instead would remain a subsonic `breeze' or could even collapse into an inflow. One way to assess whether the atmosphere of a close-in giant would be escaping in the form of a wind (i.e., becoming transonic) or not, is to find out whether there is a supersonic shocked outflow, i.e., the shock position is above the planetary outflow sonic point $r_s$. The flow beyond the sonic point cannot affect the flow inside the sonic point as information cannot be passed back through the sonic point (Figure \ref{fig.pl_conditions}; \citealt{1998A&A...330L..13D}). Once the shock is pushed to the critical point, the lower part of the atmosphere can `communicate' with the external medium: the pressure at the base of the atmosphere becomes small and the flow might reverse into an inflow.

In our work, we first derive the position of the shock, by assuming ram pressure balance between the planetary outflow and the stellar wind.\footnote{Although this is a reasonable approximation for calculating the stand-off distance to the shock, it neglects that the shocked material is heated up, thus thermal pressure inside the shock should also contribute to balance the stellar wind ram pressure \citep{2016ApJ...832..173S, 2019MNRAS.489.5784C}. In other words, our calculation neglects the thickness of the shock.} Once the shock position is known, we then need to verify whether it is above the sonic point of the planetary outflow, so that it is in the regime shown in Figure \ref{fig.pl_conditions}a, or if the shock is below the `would be' sonic point of the planetary outflow, as shown in Figure \ref{fig.pl_conditions}b. In other words, the ram pressure at the sonic point is the crucial parameter, which determines whether or not the exoplanetary atmosphere is confined by the stellar wind. 

\begin{description}
\item {\it If the ram pressure of the stellar wind at the orbital distance, $P_{\rm ram, sw} (a_{\rm orb})$, is larger than the ram pressure of the exoplanetary wind at the sonic point,  $P_{\rm ram, pw} (r_s)$, then the planetary atmospheric escape rate will be affected by the stellar wind. } 
\end{description}

Mathematically, this is written as
\begin{eqnarray}
P_{\rm ram, sw} (R=a_{\rm orb}) \lesssim P_{\rm ram, pw} (r=r_s) ~~~ \to ~~~ {\rm Fig.~\ref{fig.pl_conditions}a, ~unconf }  \\ 
P_{\rm ram, sw} (R=a_{\rm orb}) \gtrsim P_{\rm ram, pw} (r=r_s) ~~~ \to ~~~ {\rm Fig.~\ref{fig.pl_conditions}b, ~ confined.}  
\end{eqnarray}
We used the subscripts `sw' and `pw' to refer to variables computed for the stellar wind and planetary wind (i.e., escaping atmosphere), respectively, to avoid confusion. The ram pressure exerted by the planetary outflow at the sonic point $r=r_s$ is
\begin{equation}\label{eq.pram}
P_{\rm ram, pw} (r_s) = \rho(r_s) u(r_s)^2 =\gamma \frac{  \rho(r_s)  k_BT(r_s)} {m(r_s)}= \gamma P(r_s),  
\end{equation}
where we used the fact that, at the sonic point, the outflow velocity is $u(r_s)=(\gamma k_B T/m)^{1/2}$. 
For planets orbiting at close distances to their host stars, the planet's orbital Keplerian velocity  $u_K= (GM_\star/a_{\rm orb})^{1/2}$ might be comparable to or even larger than the local stellar wind velocity \citep[e.g.][]{2010ApJ...722L.168V}. In these cases, the ram pressure exerted by the stellar wind on the planet is given by $P_{\rm ram, sw} (a_{\rm orb}) = \rho\sw(a_{\rm orb}) [{\bf u}\sw(a_{\rm orb}) - {\bf u_K}]^2 $, where ${\bf u}\sw - {\bf u_K}$ is the relative velocity of the planet through the stellar wind. Given that the stellar wind velocity only has a radial component and that the Keplerian velocity (assuming circular orbit) only has an azimuthal component, we can write
\begin{equation}\label{eq.pram}
P_{\rm ram, sw} (a_{\rm orb}) = \rho\sw(a_{\rm orb}) \left[u\sw(a_{\rm orb})^2 + \frac{GM_\star}{a_{\rm orb}} \right].
\end{equation}
Therefore, using the previous expressions, our condition for confinement is
\begin{eqnarray}
\frac{ \rho\sw(a_{\rm orb}) }{\gamma P(r_s) } \left[u\sw(a_{\rm orb})^2 + \frac{GM_\star}{a_{\rm orb}} \right] \lesssim 1 ~~ \to ~~ {\rm Fig.~\ref{fig.pl_conditions}a, ~unconf.}  \label{eq.non_confinement} \\
\frac{ \rho\sw(a_{\rm orb}) }{\gamma P(r_s) } \left[u\sw(a_{\rm orb})^2 + \frac{GM_\star}{a_{\rm orb}} \right] \gtrsim 1~~ \to ~~ {\rm Fig.~\ref{fig.pl_conditions}b, ~ confined.}   \label{eq.confinement}
\end{eqnarray}

In the calculations we present in this paper, we compute two independent models: one for the planetary outflow and one for the stellar wind. With this setup we are able to evaluate the conditions shown in Equations (\ref{eq.non_confinement}) and (\ref{eq.confinement}), but we are not able to  evaluate the hydrodynamic effects that the stellar wind has on the escaping planetary atmosphere. This is currently being done in a forthcoming work, using 3D hydrodynamic simulations of interacting planetary and stellar winds.

%%%%%%%%%%%%%%%%%%%%%%%%%%%%%%%%%%%%%%%%%
\subsection{Confined vs non-confined regimes}
\begin{figure*}
	\includegraphics[width=.49\textwidth]{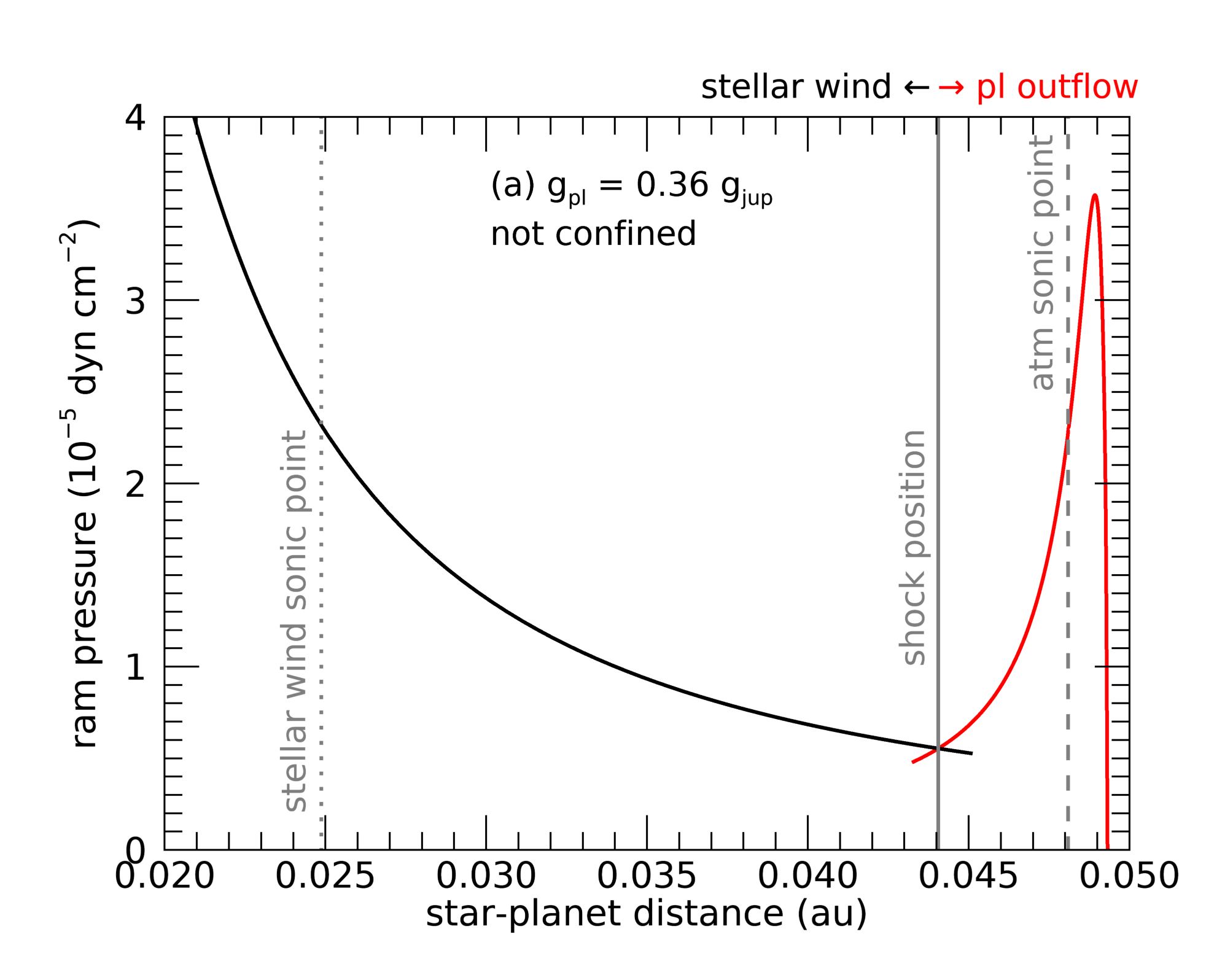}
	\includegraphics[width=.49\textwidth]{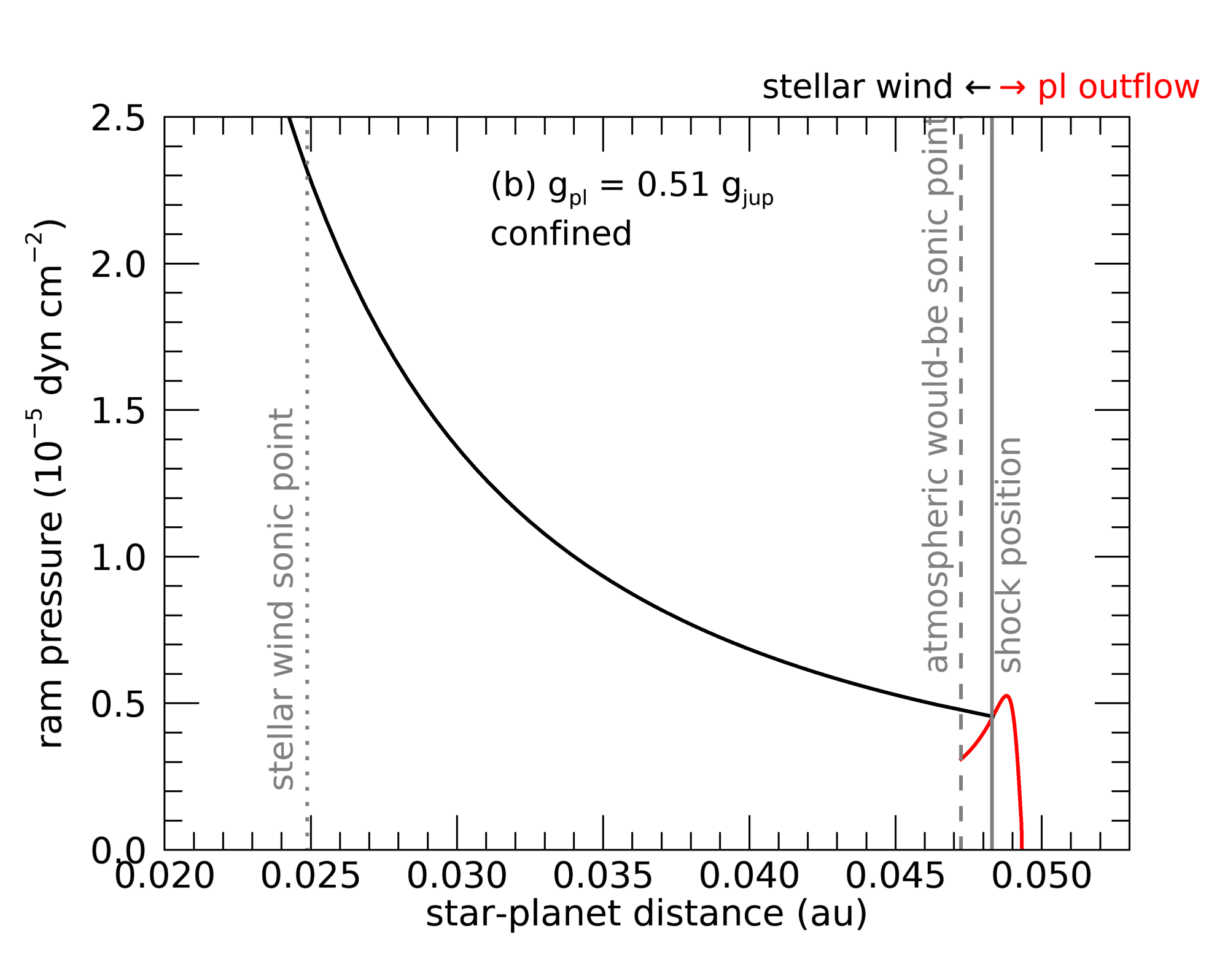}
    \caption{Ram pressure profiles for the stellar (black) and planetary (red) outflows, along the star-planet line.  The planet is assumed to orbit at $0.05$~au, and its atmosphere is expanding towards the star (located on the left of the $x$ axes, not shown in the figure). The solid vertical lines show the position of the shock, where there is pressure balance between the two flows. The sonic point of the stellar wind is indicated by the dotted line and the dashed line indicates the sonic point of the escaping atmosphere. Left: The regime where the planetary outflow is not confined by the stellar wind occurs where the shock position is above the sonic point of the planetary atmosphere (Figure \ref{fig.pl_conditions}a). Here, we assume a planetary gravity that is $36\%$ of Jupiter's surface gravity. Right: The regime where the planetary outflow is confined by the stellar wind occurs where the shock position is further down into the planetary atmosphere, below what would have been the sonic point of the planetary escaping atmosphere (Figure \ref{fig.pl_conditions}b). Here, we assume a planetary gravity that is $51\%$ of Jupiter's surface gravity.}
    \label{fig.pl_conditions2}
\end{figure*}

To exemplify the two different regimes of atmospheric confinement, we show in Figure \ref{fig.pl_conditions2} the ram pressure for the stellar wind (black) and for the planetary outflow (red), along the star-planet line. The planet is assumed to orbit at $0.05$~au, and its atmosphere is expanding towards the star (located on the left of the $x$ axes, not shown in the figure). Thus, the atmospheric outflow (red line) starts at $0.05$~au and expands towards smaller values in the $x$-axis. Figure \ref{fig.pl_conditions2}a considers a planet with radius $1.4 R_{\rm jup}$ and  mass $0.7 M_{\rm jup}$, thus having a gravity that is $36\%$ of Jupiter's surface gravity ($g_{\rm jup}$). This low gravity means that escape is easier to occur in this planet, thus its atmosphere can be accelerated to higher velocities. This, in turn, leads to ram pressures that are larger than that of the planet shown in panel b, which has the same radius, but a larger mass of  $1 M_{\rm jup}$ and a larger gravity ($51\%$ of Jupiter's surface gravity). The stellar wind in both cases are the same -- we adopt here the wind of a star of 4.6~Gyr that is similar to the present-day solar wind (Table \ref{tab.inputs}). The stellar wind is supersonic at the position where it shocks with the planetary atmosphere; the shock position is marked with vertical solid lines in both panels. At this position, we find similar local stellar wind conditions for both cases. The local stellar wind ram pressure is $0.55 \times 10^{-5}$ dyn cm$^{-2}$, with a local stellar wind temperature of $9.5\times 10^5$~K and number density of $10^4$~cm$^{-3}$ for the case plotted in Figure \ref{fig.pl_conditions2}a. For the case plotted in Figure \ref{fig.pl_conditions2}b, the ram pressure at the shock position is $0.45 \times 10^{-5}$ dyn cm$^{-2}$, with a local stellar wind temperature of $9.4 \times 10^5$~K and number density of $0.9 \times 10^4$~cm$^{-3}$. The difference between the non-confined case (panel a) and the confined case (panel b) lies at the relative position between the sonic point of the escaping atmosphere (dashed line) and the shock position (vertical line). In the first case, the shock occurs above the atmospheric sonic point, similar to the sketch shown in Figure \ref{fig.pl_conditions}a (remember that the atmosphere starts to escape from 0.05~au, expanding towards smaller values of $x$-axis, so the position of the shock occurs at smaller orbital distances than the sonic point). In the second case, the shock occurs further down the planetary atmosphere, in a region that is still subsonic (i.e., below the `would-be' sonic point), similar to the illustration in Figure \ref{fig.pl_conditions}b. 

%%%%%%%%%%%%%%%%%%%%%%%%%%%%%%%%%%%%%%%%%%%%%%%%%%%%%%%%%%%%
%
\section{Parametric study}\label{sec.parametric_results}
To investigate the regimes where stellar winds can confine atmospheric escape, we perform a parametric study, varying the system's age, planetary gravity and orbital distance. Here, we assume planetary systems at 4 different ages: 1, 2.7, 4.6 (solar age) and 6.9~Gyr. For each of the 4 ages,  we assign a stellar rotation period (from Equation \ref{eq.skumanich}) and an EUV luminosity  \citep[from the $L_{\rm EUV}$-age relations of][]{2015A&A...577L...3T}, the latter of which is used to calculate the input EUV flux for the planetary outflow calculation. From the stellar rotation period, we then assign a stellar wind base density (Equation \ref{eq.basedens}) and temperature (Equations \ref{eq.baseT_A} or  \ref{eq.baseT_B}), so that we can calculate the stellar wind models. Table \ref{tab.sw} shows the stellar wind mass-loss rates, their range of ram pressures, densities and temperatures calculated from 0.04 to 0.14~au, which are the range of orbital distances adopted in the our parametric study. 

\begin{table*}
\caption{{Stellar wind mass-loss rates, ram pressures, total number densities and temperatures calculated within $[0.04,\, 0.14]$~au. Superscripts A and B indicate different stellar wind models (Equations (\ref{eq.baseT_A}) and  (\ref{eq.baseT_B})).}}\label{tab.sw}
\begin{tabular}{ccccccccc}
\hline 
age   & ${\dot{M}\sw}^A/10^{-14}$ & $P_{\rm ram, sw}^A$/$10^{-7}$ & $n_A$ & $T_A$ & ${\dot{M}\sw}^B/10^{-14} $ & $P_{\rm ram, sw}^B$/$10^{-7}$ & $n_B$ & $T_B$ \\ 
(Gyr)&($ M_\odot/$yr) &(dyn~cm$^{-2}$) & ($10^{4}$ cm$^{-3}$) & ($10^5$K)& ($M_\odot/$yr) &(dyn~cm$^{-2}$) & ($10^{4}$ cm$^{-3}$) & ($10^5$K) \\\hline \hline
1.0& $69$ & $[4500, \,409]$ & $[29, \,1.6]$ & $[19, \,17]$ & $4.9$ & $[220,\,19]$ & $[3.9, \,0.19]$ & $[11, \,9.2]$ \\
2.7& $17$ & $[910, \,83]$ & $[9.7, \,0.49]$ & $[14, \,12]$ & $2.4$ & $[100,\,8.6]$ & $[2.0, \,0.093]$ & $[10, \,8.6]$ \\
4.6& $1.6$ & $[69, \,5.6]$ & $[1.4, \,0.064]$ & $[9.6, \,8.2]$ & $1.6$ & $[69,\,5.6]$ & $[1.4, \,0.064]$ & $[9.6, \,8.2]$ \\
6.9& $0.078$ & $[3.4, \,0.21]$ & $[0.12, \,0.0044]$ & $[6.7, \,5.7]$ & $1.2$ & $[50,\,3.9]$ & $[1.1, \,0.047]$ & $[9.4, \,8.0]$ \\
\hline
\end{tabular}
\end{table*}

Our planetary outflows are calculated for surface gravities in the range $g_{\rm pl} = [0.36, \,0.87] g_{\rm jup}$. In practice, we assume the same planetary radius of $1.4 R_{\rm jup}$ for all the atmospheric escape simulations and vary the planetary masses from $0.7$ to $1.75 M_{\rm jup}$.\footnote{We demonstrate in  Appendix \ref{sec.appendix_gravity} that interchanging $\{M_{\rm pl}, \,R_{\rm pl}\}$ by planetary gravity is a reasonable approximation for calculating planetary outflows, thus we combine these two input parameters $\{M_{\rm pl}, \,R_{\rm pl}\}$ into one input parameter $g_{\rm pl}$.} In summary, for each of the 4 ages, there are two stellar wind models (A and B), and between 150 to 220 planetary outflow models, for a variety of gravities ($g_{\rm pl} = [0.36, \,0.87]~ g_{\rm jup}$) and orbital distances ($[0.04,\, 0.14]$~au). 

For all these planetary outflow simulations, we calculate the Knudsen number $K_n={\lambda_{\rm mfp}}/{h}$  to check that the atmosphere remains collisional ($K_n \ll 1$). Here, $h=\rho/|d\rho/dr|$ is the scale height of the atmosphere and $\lambda_{\rm mfp} $ is the mean-free path. Closer to the planet, the high densities ensure that  the outflow is collisional. However, at higher altitudes, the densities can become very low, and the atmosphere can become collisionless, in which case particles would travel in ballistic trajectories (not following a Maxwellian distribution). In our simulations, we confirm that the collisional approximation holds by calculating Knudsen numbers at the Roche lobe distance (the upper boundary in our simulations), for the ionised and for the neutral flows. For the ionised flow, we consider  proton-proton scattering, where the mean-free path  of a proton is $\lambda_{\rm mfp} = (n_p \sigma_{c})^{-1}$, where  $\sigma_{c} = 10^{-13} (T/10^4)^{-2}$~cm$^{2}$ is the Coulomb's cross section for proton-proton scattering and $n_p$ the proton number density. In this case, our Knudsen numbers range from $K_{n,c} (R_{\rm Roche})= 10^{-6}$ up to $\sim 3\times 10^{-4}$ for Coulomb collisions. In addition to Coulomb collisions, collisions between two hydrogens (proton-neutral or neutral-neutral) can also be computed. Usually, in  atmospheric models of close-in giants, a typical value adopted for neutral-neutral or proton-neutral (i.e., charge-exchange) cross section is $\sigma_{H} \simeq 3.3\times 10^{-15}$ cm$^2$  (\citealt{2005ApJ...621.1049T,2007P&SS...55.1426G, 2016A&A...586A..75S}), and hard-sphere collisions are less important \citep{2011ApJ...733...98G}. In this case, the mean-free path of collisions between  hydrogens become  $\lambda_{\rm mfp} = (n_H \sigma_{H})^{-1}$, where $n_H$ is the total hydrogen number density. Similarly to the calculation done for Coulomb collisions, we calculate the Knudsen number for hydrogen collisions at  the Roche lobe, and find $K_{n,H} (R_{\rm Roche}) \sim 10^{-4}$ up to $2.6$ (the average over all simulations is 0.24). The systems in which $K_{n,H} \gtrsim 1$ at the Roche lobe (39 out of 719 simulations) are the ones with large surface gravities and low EUV fluxes (larger orbital radii). These are also the systems with Roche lobes extending the farthest from the planet (thus, local densities are lower).  As we will see below, these are also the planets whose atmospheres are believed to be confined by the stellar wind and thus their atmospheres are unlikely to extend out to the Roche lobe, indicating that even the neutral atmosphere remains collisional in the cases studied here.

\begin{figure*}
	\includegraphics[width=.49\textwidth]{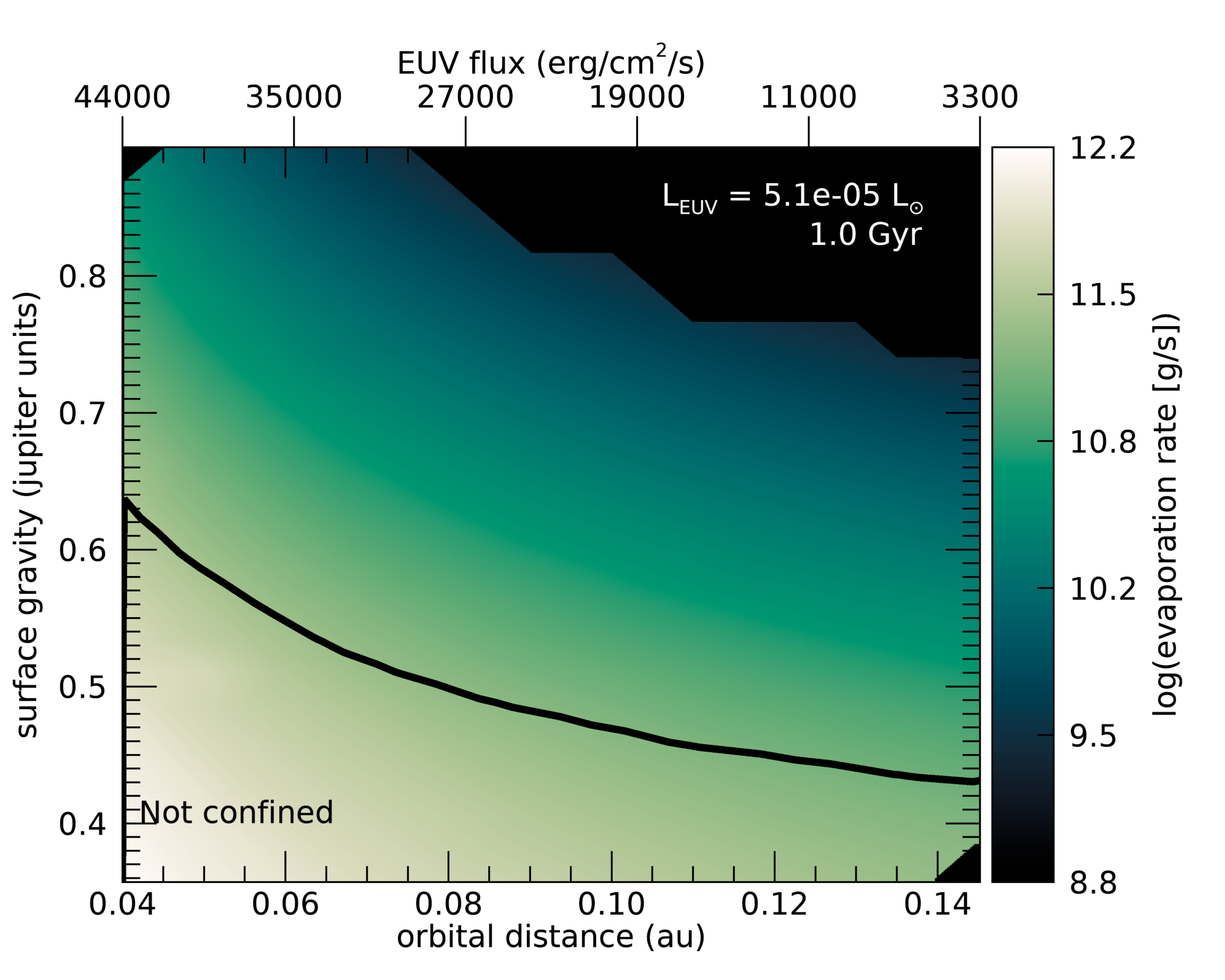}
	\includegraphics[width=.49\textwidth]{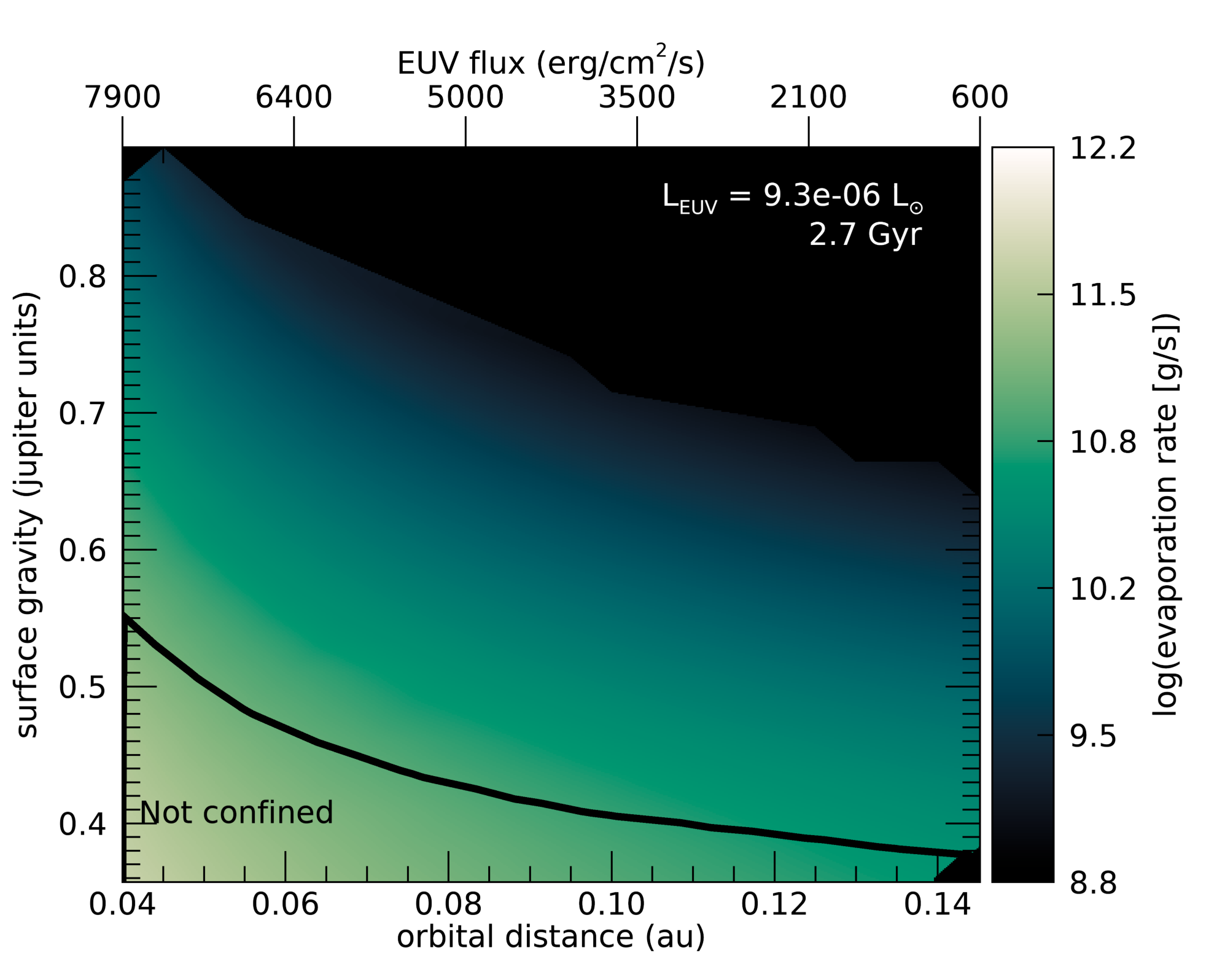}\\
	\includegraphics[width=.49\textwidth]{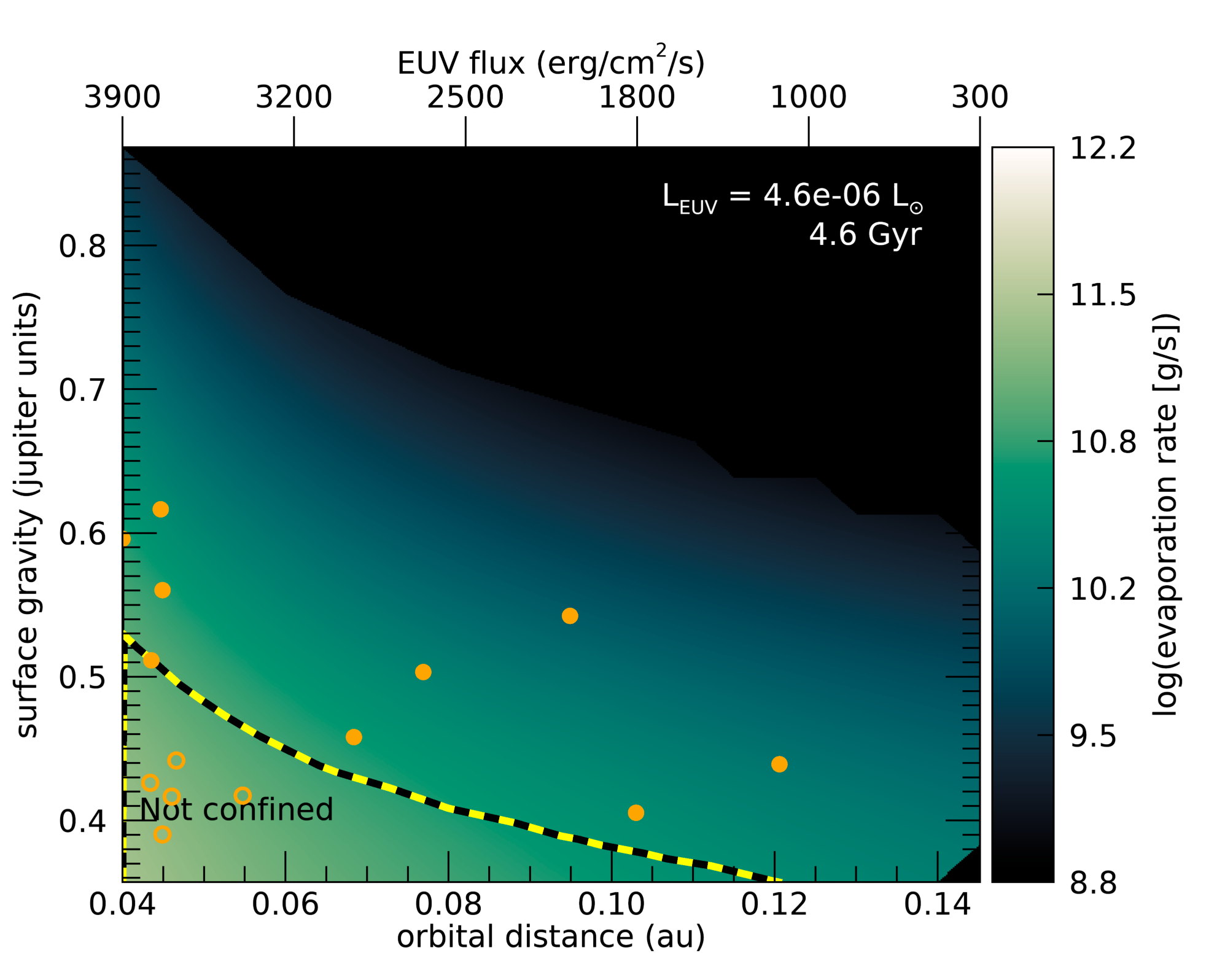}
	\includegraphics[width=.49\textwidth]{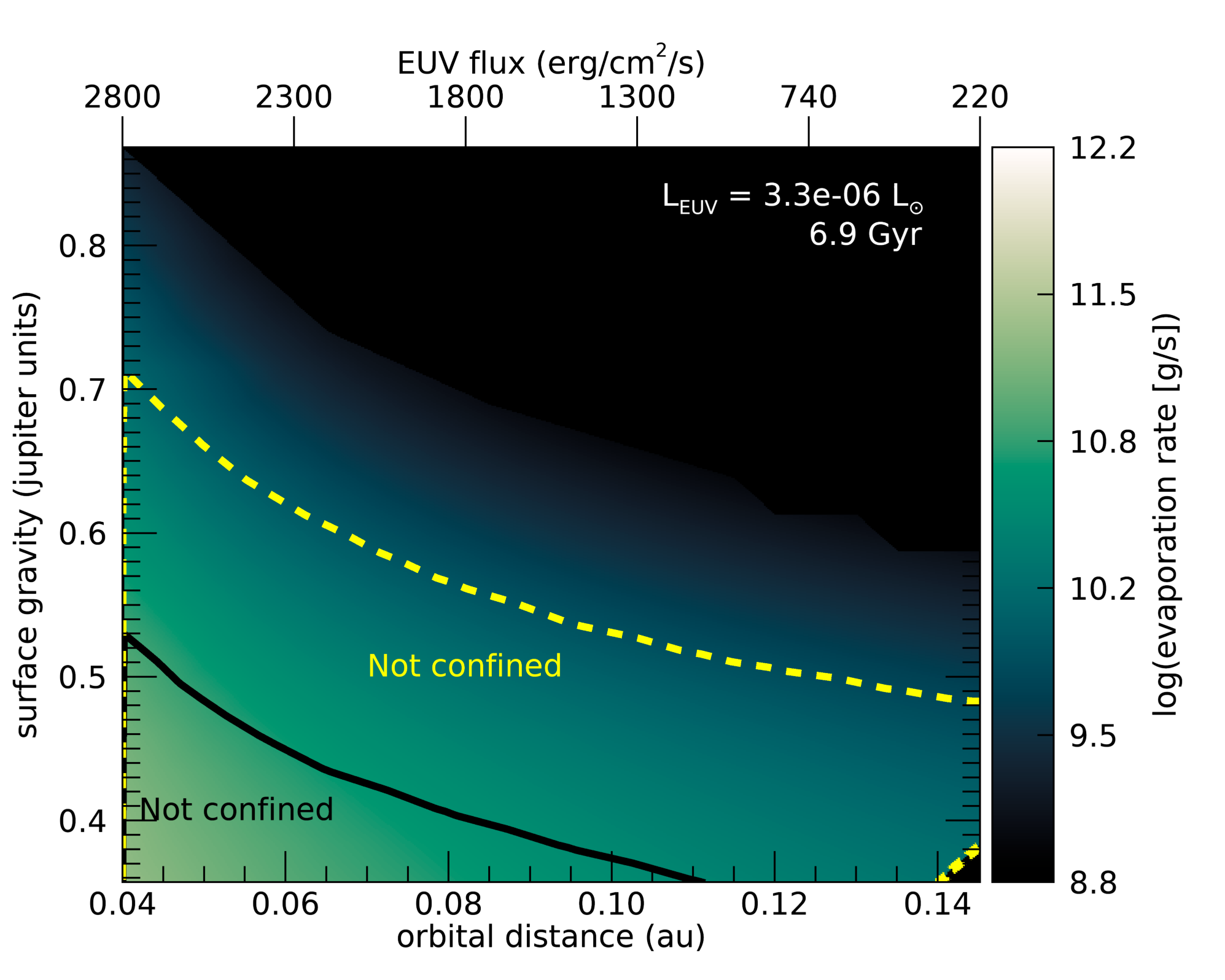}
    \caption{Planetary evaporation rates (colour) as a function of surface gravity ($y$ axis) and orbital distances ($x$ axis), for planets orbiting stars of different ages, from young (top left) to older (bottom right) stars.  The associated EUV luminosity of each star, along with its age, are shown on each panel. An EUV flux scaling in shown in the top $x$ axis for reference. The black lines show the different regimes where planetary outflow will not be confined (below black lines, Figure \ref{fig.pl_conditions}a), in which case the planetary outflow rate should not be modified by the stellar wind, or will be confined (above black lines, Figure \ref{fig.pl_conditions}b). In the latter case, planetary escape models that do not take into account the presence of the stellar wind might be overestimating atmospheric escape rates (including our models above the black lines). The black lines assume Model B for the stellar wind, while the yellow dashed lines assume Model A. The yellow dashed lines do not appear in the top two panels, as according to this stellar wind model, planetary escape would be fully confined in the studied parameter space. Black and yellow lines are the same at the solar age (4.6 Gyr), as wind Models A and B are identical at this age by construction. The circles shown in this panel refers to 14 known exoplanets (see Section \ref{sec.data}). For the oldest star, the region of non-confinement is larger for Model A (yellow), as the wind is weaker than that computed with Model B.}
    \label{fig.mdot_param}
\end{figure*}

Figure \ref{fig.mdot_param} shows the evaporation rate of the planetary atmosphere as a function of orbital distance and surface gravity. Each panel indicates a different age and, thus, a different $L_{\rm EUV}$. For easy reference, a conversion from orbital distance to EUV incident flux is shown in the top $x$ axis of each panel. We notice from these panels that evaporation rates are larger for closer-in planets and lower gravity planets, similar to what has been seen in \citet{2018A&A...619A.151K, 2019MNRAS.490.3760A}. Additionally, because  younger stars have higher $L_{\rm EUV}$, at a same surface gravity and orbital distance, planets have higher escape rates when orbiting younger stars. Note that the upper right portions of these panels do not have computed models\footnote{Because the planet does not receive enough energy flux at larger $a_{\rm orb}$, fewer models converge at higher orbital distances. Similarly, the higher surface gravities means it is more difficult to lift atmospheric material, and thus less of our models converge at higher $g_{\rm pl}$.} and we saturate the minimum evaporation rate to $10^{8.8}$~g/s.

The solid black line and yellow dashed lines in Figure \ref{fig.mdot_param} separate the two regimes discussed in the previous Section. We will first concentrate on the black lines, which are related to the stellar wind Model B. Below the black line, the ram pressure of the stellar wind is smaller than the ram pressure of the planetary outflow at the sonic point (Equation \ref{eq.non_confinement} and  Figure~\ref{fig.pl_conditions}a), hence the atmosphere is `not-confined'. In other words, planetary outflow rates should not be modified by the stellar wind, although the stellar wind is expected to shape the outflow \citep[see, e.g., 3D simulations of interacting flows][]{2018MNRAS.479.3115V, 2019ApJ...873...89M}. We also see in Figure \ref{fig.mdot_param} that the non-confined region decreases from 1~Gyr to 2.7~Gyr and then to 4.6~Gyr for Model B, but does not change significantly from 4.6 to 6.9~Gyr. 

Conversely, above the black line, the stellar wind ram pressure is larger than the ram pressure of the escaping planetary atmosphere at the sonic point and thus the stellar wind can reduce or even prevent atmospheric escape in these regions of the diagrams. {\it Planetary escape models that do not take into account the presence of the stellar wind might be overestimating escape rates in these regions.} This is also the case of the planetary outflow models we use in this paper, since they do not account for the presence of the stellar wind. Recent multi-dimensional simulations have investigated how stellar winds can affect planetary outflows \citep{2014MNRAS.438.1654V,2018MNRAS.479.3115V,2017MNRAS.466.2458C, 2019ApJ...873...89M, 2019MNRAS.483.1481D, 2019ApJ...885...67K}. To the best of our knowledge, the large majority of these simulations were in the `not confined' scenario, and did not consider a situation in which the atmosphere of the planet does not accelerate above sonic speed (for an exception, see e.g., \citealt{2016ApJ...820....3C}). 

%%%%%%%%%%%%%%%%%%%%%%%%%%%%%%%%%%%%%%%%%%%%%%%%%%%%%%%
\subsection{Effects of different assumptions for the stellar wind}
The yellow dashed lines  in Figure \ref{fig.mdot_param} play the same role as the black lines, except that they refer to Model A. Model A is the stellar wind model that assumes a temperature break for slowly rotating stars (in our simulations, only the 1~Gyr case falls in the `fast rotating branch', with $P_{\rm rot, \star} \lesssim 0.7 P_{\rm rot, \odot} $). Regardless of the stellar wind model, there is always a region in our parameter space where the planet escape is  `confined', i.e., the stellar wind pressure is high enough to prevent the planetary outflow to expand beyond sonic velocities. Within our parameter space (i.e., $a_{\rm orb} = [0.04,\, 0.14]$~au, $g_{\rm pl} = [0.36, \,0.87]~ g_{\rm jup}$), the confined region preferably occurs at higher planetary gravities and higher orbital distances. 

Comparing the stellar wind mass-loss rates and ram pressures presented in Table \ref{tab.sw}, we see that Models A and B have the same wind characteristics at solar age  (4.6 Gyr panel in Figure \ref{fig.mdot_param}) -- this is by construction, since these two models are anchored in solar wind values. For younger stars, Model A has higher mass-loss rates and local ram pressures than Model B, due to Model A's larger temperature. As a consequence, the local ram pressures of Model A for 1 and 2.7 Gyr are much higher than the ram pressures of the escaping atmosphere. Thus, the yellow dashed lines do not appear in the top two panels of  Figure \ref{fig.mdot_param}, as according to this stellar wind model, planetary escape would be fully confined in the studied parameter space. For the oldest star (6.9 Gyr), the situation is reversed: the wind predicted by Model A is now weaker than that predicted by Model B. Thus, the region of non-confinement is larger for Model A (yellow), as the wind is weaker than that computed with Model B. 

This simple exercise shows us that it is important to know the stellar wind conditions in order to predict the evaporating conditions of a close-in planet. If we detect an evaporating planet, we will know that the stellar wind would not have been able to confine its escape. Thus, detecting evaporation of a planet could also help us constrain somewhat the stellar wind properties (\citealt{2017MNRAS.470.4026V, 2020arXiv200308812M}, Villarreal D'Angelo et al, in prep).

%%%%%%%%%%%%%%%%%%%%%%%%%%%%%%%%%%%%%%%%%%%%%%%%%%%%%%%
\subsection{Are known exoplanets confined by the winds of their hosts?}\label{sec.data}
Most of the parameter space studied here lies in the `confined' region. To get a rough idea of how many known planets would fall in our range of studied parameters, we compiled a list of exoplanets with  $a_{\rm orb} = [0.04,\, 0.14]$~au and $g_{\rm pl} = [0.36, \,0.87]~ g_{\rm jup}$ using data from the NASA Exoplanet Catalogue (Dec 6th, 2019). We restricted the mass of the host star to be $1\pm 0.1~M_\odot$ and its radius $1\pm 0.1~R_\odot$. We also limited the systems that were within 500~pc. From this, we ended up with 14 exoplanets. Given that we do not know the ages of these systems, we assumed that these are all old-ish stars and overplotted them to the 4.6~Gyr-panel of Figure \ref{fig.mdot_param}. This assumption is  not unreasonable as current planet detection methods favour older systems.\footnote{Notice that planets like WASP-12b and GJ\,436b, which have been observed to undergo strong evaporation \citep{2010ApJ...714L.222F,2017A&A...605L...7L}, did not meet our filtering criteria. In the case of WASP-12b, its orbital distance is smaller than the minimum orbital distance considered here. In the case of GJ\,436b, its host star is an M dwarf, while we focused here on solar-mass stars.}

There were 5 objects that fell in the `not-confined' regime (empty orange circles): HAT-P-55b, HAT-P-28b, WASP-124b, HATS-29b and HAT-P-25b. These are the planets that would show higher escape rates and whose evaporation rates would not be affected by the stellar wind, according to our models. The remaining 9 planets in our sample fell in the `confined' regime, they are (filled orange circles): Kepler-1655b, Kepler-20c, Kepler-447b, WASP-28b, $\pi$ Men c, HAT-P-27b, HATS-10b, HATS-30b, K2-287b. According to our model, even though the atmospheres of these planets are highly irradiated, they would be bound by an external stellar wind pressure which would affect their escape rates. 

In particular, our prediction of a reduced atmospheric escape in $\pi$ Men c, a {highly irradiated sub-Neptune}, is in line with the  non-detection of hydrogen escape reported by \citet{2019arXiv191206913G}. The expectation was that escape rate in $\pi$ Men c would be comparable or even larger than that of GJ\,436b, a warm-Neptune that shows one of the most impressive neutral hydrogen transits ever detected \citep{2014ApJ...786..132K,2017A&A...605L...7L}. In their work, \citet{2019arXiv191206913G} attributed this surprising non-detection of escaping hydrogen in $\pi$ Men c due to its atmospheric composition, which would not be hydrogen dominated, as opposed to GJ\,436b, which would have a hydrogen-dominated atmosphere. If the atmosphere of $\pi$ Men c is non-hydrogen-dominated, these authors still expect  high evaporation rates of other chemical species, which could be detected in spectroscopic transits of metal lines, such as OI and CII. 
They thus proposed that these two different behaviours in atmospheric composition could be tested by searching  for transits in metal lines, in addition to neutral hydrogen transits. If, conversely, the explanation of the non-detection of hydrogen escape in  $\pi$ Men c  is due to a stellar wind confinement as we predict here, our model would  predict lack/reduced escape in {\it any} chemical species, i.e., the reduced escape would not depend on the atmospheric composition. Observing metal line transits of $\pi$ Men c could be one way to test between these different interpretations for the lack of hydrogen escape in $\pi$ Men c.

%74 objects if only filtering stellar mass, radius and distance. adding orbital distance and gravity constraint, then we get 14 planets

%filter=where(pl_orbsmax ge 0.04 and pl_orbsmax le 0.14 and grav ge 0.35 and grav lt 0.88 $
%  and st_mass ge 0.9 and st_mass le 1.1 and st_rad ge 0.9 and st_rad le 1.1 and st_dist lt 500 )

%, thus, according  to stellar wind Model A, close-in giants within 0.04 to 0.14~au
%
%%HAT-P-55 b   0       480.00000
%HAT-P-28 b   0       395.00000
%WASP-124 b   0       433.00000
%HATS-29 b   0       351.00000
%HAT-P-25 b   0       302.95000
%Kepler-1655 b   1       214.58000
%Kepler-20 c   1       284.88000
%Kepler-447 b   1       270.12000
%WASP-28 b   1       410.00000
%pi Men c   1       18.280000
%HAT-P-27 b   1       204.00000
%HATS-10 b   1       496.00000
%HATS-30 b   1       339.00000
%K2-287 b   1       159.05000

%%%%%%%%%%%%%%%%%%%%%%%%%%%%%%%%%%%%%%%%%%%%%%%%%%%%%%%%%%%
\section{Discussion of model limitations}\label{sec.discussion}

%%%%%%%%%%%%%%%%%%%%%%%%%%%%%%%%%%%%%%%%%%%%%%%%%%%%%%%%%%%
\subsection{Inhomogeneities in the stellar wind and the stability of escaping atmosphere}
In this work, we assume a spherically symmetric stellar wind in steady state. Thus, stellar wind properties would not change along the orbital path of a planet in circular orbit. However, due to complex surface magnetic field geometries, it is more likely that stellar winds show inhomogeneities, such as for example, streamers with high velocity winds, and density variations with longitude/latitude \citep[e.g.][]{2011MNRAS.414.1573V,2014MNRAS.438.1162V}. Additionally, short-term events such as coronal mass ejections and flares can also affect planetary escape \citep{2017ApJ...846...31C,2018ApJ...869..108B}. Therefore, it is possible that the variation in stellar outflow properties and stellar energy input could induce the planet to move from different confinement regimes, in timescales that could be smaller than one planetary year. This could mean that, at certain orbital phases, the planetary evaporation rate could suffer an abrupt drop-off, when going from a regime of non-confinement (low stellar wind ram pressure) to confinement (high stellar wind ram pressure). Would escape rates then turn on again, once the planet moves back to a  non-confinement regime? 

To answer this question, one needs to investigate the stability of the atmospheric escape solution, which is not dealt with in the present paper. In general, supersonic shocked flows (like in the non-confinement regime) are stable \citep{1998A&A...330L..13D}. On the contrary, the subsonic flows in the confined regime might not be stable. \citet{1994ApJ...432L..55V} suggested that the flow might generate a hysteresis-type cycle, in which going from supersonic to subsonic, the flow changes from a `wind' to a `breeze', but coming back from subsonic to supersonic does not mean that the flow changes back from `breeze' to `wind'. Instead, the flow can reverse into a supersonic accretion, which is stable, but no longer is an outflow. Therefore, the equilibrium of the flow does not only depend on the sub or supersonic nature of the flow, but also on its previous history \citep{1998A&A...330L..13D}.

%%%%%%%%%%%%%%%%%%%%%%%%%%%%%%%%%%%%%%%%%%%%%
\subsection{Effects of magnetic fields}

Throughout the paper we have assumed the cases of unmagnetised planets and winds. Stellar magnetic fields can create `dead zones' that do not contribute to stellar mass-loss. The stellar wind plasma remains trapped within the dead zone, whose size is determined by the strength of the magnetic field and plasma thermal properties  \citep{2009ApJ...699..441V}. Additionally, stellar magnetism shapes stellar winds, converting winds from a spherically symmetric outflow to outflows with more complex velocity and density distributions. This, in turn, affects the environment around planets, creating the inhomogeneities we discussed in the previous subsection.  Stellar magnetism also alters the external ambient pressure around planets, as it contributes to pressure confinement. In their parametric study, \citet{2014MNRAS.444.3761O} demonstrated  that, as the  stellar magnetic field strength is increased, and, thus, the external ambient pressure around an evaporating exoplanet, escape rates could either be enhanced {or} suppressed. Thus, active and moderately-active stars, with their intense winds and magnetism, could actually reduce or even suppress atmospheric loss from their exoplanets, instead of increasing atmospheric erosion.

Similarly, dead-zones can also be generated in case of magnetised planetary outflows \citep{2015ApJ...813...50K}. Magnetic fields thus are expected to alter escape rates of close-in giant planets and, even in weakly magnetised planets,  magnetism affects atmospheric loss processes, and thus escape rates \citep{2001Sci...291.1939S,2007SSRv..129..245L, 2018GeoRL..45.9336S, 2019MNRAS.488.2108E}.

Altogether, the combined effects of stellar {\it and} planetary magnetic fields is not immediate to grasp and, among other factors, it would depend on the details of the geometries of such fields. For example, even in the simplest scenario where the star and the planet possess (anti-)aligned dipolar fields, the relative orientation of such fields can generate `closed' or `open' planetary magnetospheres  \citep {2019MNRAS.489.5784C, 2019ApJ...877...80B, 2019ARep...63..550Z}, which would affect outflow rates. To quantify the effects of stellar and magnetic fields in atmospheric confinement, we need to switch from the simple 1D models to multi-dimensional simulations.

%%%%%%%%%%%%%%%%%%%%%%%%%%%%%%%%%%%%%%%%%%%%%%%%%%%%%%%%%%%
\subsection{Effects on the nightside}
Another limitation of 1D escape models is that they only consider escape along the planet-star line (known as the dayside), where the stellar irradiation is maximum. On the diametrically opposite side of the planet, in the nightside, the lack of irradiation would likely reduce atmospheric escape there. It is  possible that meridional atmospheric winds would redistribute the heat from the dayside to the nightside. In that case, the nightside could still present some evaporation. 

An additional characteristic not captured in 1D models is that, due to high orbital speeds of close-in planets, the shock interface is formed at an angle with the orbital motion, while stellar irradiation always impact on the dayside of the planet. This means that the point in the planet where the irradiation effect is maximum does not coincide with the region where the stellar wind confinement is maximum. This strong mismatch in orientations is unprecedented in solar system planets. We do not know the consequences this could have on planetary outflows and how this could modify the planetary outflow structure and, consequently, its observational signature. For example, even if we were in the scenario of `confined' escape in one region of the planet, another region could still have evaporation unconfined by the stellar wind. 

%%%%%%%%%%%%%%%%%%%%%%%%%%%%%%%%%%%%%%%%%%%%%%%%%%%%%%%%%%%%%%%%
\section{Conclusions}\label{sec.conclusion}
 In analogy to the solar wind expansion into the ISM, planetary outflows expand into the winds of their host stars, being bounded by the stellar wind external pressure. If the external pressure is sufficiently high, it can push further down the altitude at which the interaction between the two flows take place. If the interaction happens where the planetary outflow is supersonic, similar to the analogy of the ISM-solar wind interaction, the wind of the host star is expected to shape the escaping atmosphere of the planet, for example, forming an asymmetric bubble around the planet. Such an interaction would generate signatures that can be detected in observations, e.g., spectroscopic transit asymmetries, but would not affect the rate at which the atmosphere escapes. Conversely, in the limit where the interaction is pushed down below where the sonic point of the planetary outflow would have been, the stellar wind might prevent/reduce the escape of the planetary atmosphere. This happens because, in interaction with subsonic atmospheres, the lower part of the atmosphere can `communicate'  with the external medium: the pressure at the base of the atmosphere becomes small and the planetary outflow can be reduced, prevented, or even reversed into an inflow \citep{1998A&A...330L..13D, 2016ApJ...820....3C}. A sketch representing the non-confined and confined regimes, as we call them in this paper, is shown in Figure \ref{fig.pl_conditions}.

In this paper, we investigated whether atmospheric escape of close-in giants could be confined by the large pressure of stellar winds around close-in planets. For that, we modelled planetary escape in a range of close-in giants, with orbital distances $a_{\rm orb} = [0.04,\, 0.14]$~au and planetary gravities $g_{\rm pl} = [0.36, \,0.87]~ g_{\rm jup}$. Planetary escape in hot Jupiters is driven by EUV heating from the high-energy photons of the star incident on the planet. Because EUV stellar luminosities evolve with age, we considered planets orbiting stars at four different ages: 1, 2.7, 4.6 (solar age) and 6.9~Gyr. With this, we then created four different model grids of planetary escape, each represented by a different age of the system. 

As stars evolve, so do their stellar winds \citep{2019MNRAS.483..873O} and thus, for each age, we also modelled the wind of the host star using a polytropic, thermally-driven wind model. Given uncertainties in stellar wind base parameters, we adopted two stellar wind models: Model A, in which the evolution of the wind temperature is represented by a broken power law \citep[$\propto $age$^{0.19}$ for stars younger than 2.2~Gyr, and $\propto $age$^{0.6}$ for older ones;][]{2018MNRAS.476.2465O}, and Model B, in which the temperature has a shallower dependence with age \citep[$\propto $age$^{0.05}$;][]{2007A&A...463...11H}.

We showed that, regardless of the stellar wind model, there is always a region in our parameter space where the planet escape is  `confined', i.e., the stellar wind pressure is high enough to prevent the planetary outflow to expand beyond sonic velocities. Within our parameter space (i.e., $a_{\rm orb} = [0.04,\, 0.14]$~au, $g_{\rm pl} = [0.36, \,0.87]~ g_{\rm jup}$), the confined region preferably occurs at higher planetary gravities and higher orbital distances. 

Additionally, we showed that the region of our parameter space where the planet is confined or not by the stellar wind changes with age. The size of the confined region, though, depends on the stellar wind model adopted. According to Model A, in which the temperature has a strong dependence with age, the region of non-confinement only occurs for solar age and older systems. At the younger simulated ages (1 and 2.7~Gyr), our entire parameter space is considered `confined'! This is because this stellar wind model predicts quite strong stellar winds at early ages. According to Model B, regardless of the simulated age, there is always a region of non-confinement, although the size of this regions decreases with age. 

We conclude, thus, that it is important to know the stellar wind conditions in order to predict the evaporating conditions of a close-in planet. Although this might be discouraging at first sight, as we add an extra layer of uncertainty in escape models (namely, the stellar wind conditions), it can also work out in our favour: if we detect an evaporating planet, we will know that the stellar wind would not have been able to confine its escape. Thus, detecting evaporation of a planet could also help us constrain somewhat the stellar wind properties (\citealt{2017MNRAS.470.4026V, 2020arXiv200308812M}, Villarreal D'Angelo et al, in prep).

Given the fact that planetary outflows can be confined during young ages, and thus could have their escape rates reduced/halted, our results challenge  the commonly accepted scenario that evaporation rates are always higher in planets orbiting young stars. Likewise, close-in planets, especially those with higher gravities, can also have escape rates affected by the confinement of their outflows by the host star's wind, which implies that hydrodynamic escape models that do not take into account the interaction with stellar winds might be overestimating atmospheric escape rates of close-in and/or young giant planets. 

Within our parameter space, our model predicts 5 known exoplanets that would have unconfined escaping atmospheres: HAT-P-55b, HAT-P-28b, WASP-124b, HATS-29b and HAT-P-25b. Nine other known exoplanets would have their atmospheres confined and could possibly have a reduction/lack of atmospheric escape: Kepler-1655b, Kepler-20c, Kepler-447b, WASP-28b, $\pi$ Men c, HAT-P-27b, HATS-10b, HATS-30b, K2-287b. In particular, our prediction of a reduced escape in $\pi$ Men c, a {highly irradiated sub-Neptune}, is in line with a recent  non-detection of hydrogen escape \citep{2019arXiv191206913G}. This unexpected non-detection was attributed to the atmospheric composition of the planet, which would be non-hydrogen dominated. According to \citet{2019arXiv191206913G}, escape in this planet would still occur at high rates, but in heavier  chemical species, leading these authors to  suggest that escape could be probed in spectroscopic transits of metal lines, such as OI and CII.  If, conversely, the explanation of the non-detection of hydrogen escape in  $\pi$ Men c  is due to a stellar wind confinement, as we predicted here, our model would  predict lack/reduced escape in {\it any} chemical species. Thus, observing $\pi$ Men c spectroscopic transits in metal lines could be one way to test between these different interpretations for the lack of hydrogen escape in $\pi$ Men c.

%%%%%%%%%%%%%%%%%%%%%%%%%%%%%%%%%%%%%%%%%%%%%%%%%
\section*{Acknowledgements}
This project has received funding from the European Research Council (ERC) under the European Union's Horizon 2020 research and innovation programme (grant agreement No 817540, ASTROFLOW) and from the  Laidlaw Undergraduate Research and Leadership Programme at Trinity College Dublin.  This research has made use of the NASA Exoplanet Archive, which is operated by the California Institute of Technology, under contract with the National Aeronautics and Space Administration under the Exoplanet Exploration Program.

\let\mnrasl=\mnras

\bsp
\label{lastpage}

\appendix

%%%%%%%%%%%%%%%%%%%%%%%%%%%%%%%%%%%%%%%%%%%%%
\section{Combining planetary mass and radius into one model parameter: surface gravity}\label{sec.appendix_gravity}
\begin{figure}
\centering
  \includegraphics[width=0.47\textwidth]{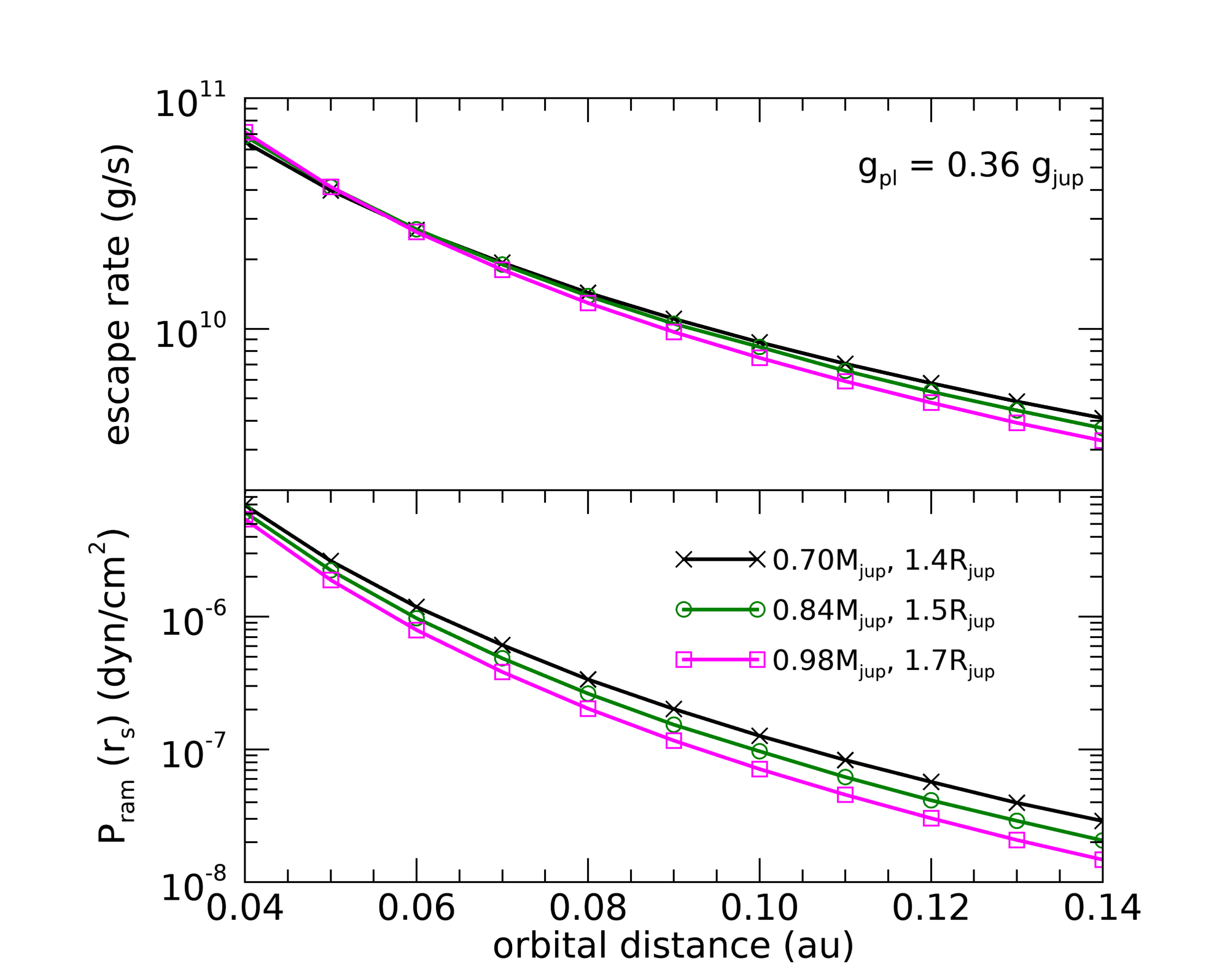}
\caption{Simulations of escaping atmospheres for planets with the same surface gravity ($0.36 g_{\rm jup}$), but different masses and radii. For each \{mass, radius\} combination, we vary the orbital distance and show the calculated evaporation rate (top) and ram pressure at the sonic radius (bottom). If the combination of two input parameters  \{mass, radius\} into a single input parameter ($g_{\rm pl}$) were perfect, we would see no difference between all these different curves. Although this is not exactly true, the relatively small differences between the three curves demonstrate that    interchanging \{mass, radius\} by planetary gravity is a reasonable approximation for calculating planetary outflows. }\label{fig:comb_param_mdot_pram}
\end{figure}

Here, we demonstrate that interchanging $\{M_{\rm pl}, \,R_{\rm pl}\}$ by planetary gravity is a reasonable approximation for calculating planetary outflows. Figure \ref{fig:comb_param_mdot_pram} shows a number of simulations for planets with the same surface gravity ($0.36 g_{\rm jup}$), but different masses and radii. For each \{mass, radius\} combination, we vary the orbital distance of the planet and assume $L_{\rm EUV} = 8.2 \times 10^{-7} L_\odot$ in all cases. The top panel in  Figure \ref{fig:comb_param_mdot_pram} shows the derived escape rate as a function of orbital distance, while the bottom panel shows the ram pressure calculated at the sonic point. Each  \{mass, radius\} combination is represented by a different curve. 

If the combination of two input parameters  \{mass, radius\} into a single input parameter ($g_{\rm pl}$) were perfect, we would see no difference between all these different curves. While this is not exactly the case, we see in  Figure \ref{fig:comb_param_mdot_pram} that these curves are not too different from each other, demonstrating thus that   interchanging $\{M_{\rm pl}, \,R_{\rm pl}\}$ by planetary gravity is a reasonable approximation for calculating planetary outflows.

\end{document}